\documentstyle[12pt,fleqn,epsfig]{article}   
\begin{document}           
\begin{center}
{\bf INSTITUT~F\"{U}R~KERNPHYSIK,~UNIVERSIT\"{A}T~FRANKFURT}\\
60486 Frankfurt, August--Euler--Strasse 6, Germany
\end{center}

\hfill IKF--HENPG/8--95
\vspace{.5cm}

\begin{center}
{  \Large \bf Strangeness  in  Nuclear Collisions
}
\end{center}

\vspace{0.3cm}
\begin{center}
{Marek Ga\'zdzicki\footnote{E--mail address:
marek@ikf.uni--frankfurt.de}
  and Dieter R\"ohrich\footnote{E--mail address:
roehrich@ikf.uni--frankfurt.de} }
\end{center}

\vspace{1cm}

\begin{center}
{\bf Abstract}
\end{center}

Data on the mean multiplicity of strange hadrons  produced
in minimum bias proton--proton and
central nucleus--nucleus collisions at momenta between
2.8 and 400 GeV/c per nucleon have been compiled.
The multiplicities  for nucleon--nucleon
interactions were constructed.
The ratios of strange particle multiplicity to participant
nucleon as well as  to pion multiplicity
are larger for central nucleus--nucleus collisions than for
nucleon--nucleon interactions at all studied energies.
The data at AGS energies suggest that the latter ratio 
saturates with increasing  masses of the colliding nuclei.
The strangeness to pion multiplicity ratio observed in  
nucleon--nucleon interactions 
increases with  collision
energy in the whole energy range studied.
A qualitatively  different behaviour is observed for central
nucleus--nucleus collisions: the ratio rapidly increases when
going from 
Dubna to AGS energies and changes little between AGS and SPS energies.
This change in the behaviour can be related to the increase
in the entropy production observed in central nucleus-nucleus 
collisions at the same energy range.
The results are interpreted within a statistical approach.
They are consistent with the hypothesis that the 
Quark Gluon Plasma is created at SPS energies, the critical 
collision energy being between AGS and SPS energies.

\vfill
\begin{center}
{\it Z. Phys.} {\bf C71} (1996) 55.
\end{center} 
\today

\newpage

\section{Introduction}

The origin of quark confinement inside hadrons and the
origin of masses of known particles are two distinct
problems of modern physics \cite{Le:95}.
It is expected that at high enough energy density 
quarks are deconfined and their masses, due to
chiral symmetry restoration, are small in comparison
with constituent quark masses or masses of hadrons.
This new form of matter conjectured long ago \cite{Co:75}
is called Quark Gluon Plasma \cite{Sh:80}.

Experimental  
studies of the Quark Gluon Plasma  properties should help to
understand the puzzles of quark confinement and particle masses.
In such investigations high energy 
nuclear collisions play a unique role \cite{QM95}.
They allow to create   `macroscopic'
(in comparison to the characteristic scale of strong interaction
$\approx$ 1 fm) space--time regions with high energy density
in the laboratory.
Their volumes and the energy densities are controlled 
by the sizes of the colliding nuclei and their collision energy.

Recent data on central Pb+Pb collisions at the highest energy available
in the laboratory  (CERN SPS: $\sqrt{s} \approx$ 20 A$\cdot$GeV)
indicate that matter with energy density of several GeV/fm$^3$
is created at the early stage of the collision \cite{Al:95a}.
This estimated energy density is significantly higher than
the critical energy density  ($\epsilon \approx$ 1 GeV/fm$^3$)
at which a transition to a
Quark Gluon Plasma takes place according to recent QCD lattice
simulations \cite{Ka:93}.
Thus the  crucial question is: Do we create a Quark Gluon Plasma
already in central nucleus--nuclues collisions at the CERN SPS?
It is difficult to answer this question 
because  
analysis of the observed potential signals of a Quark Gluon Plasma
creation, like strangeness enhancement \cite{Ko:86,Mi:88,Ga:89} and
J/$\Psi$ suppression \cite{Ma:86,Bu:88} 
is model dependent and
obscured by the problems with an exact theoretical
description of  high energy nucleus--nucleus collisions, in which
nonperturbative QCD processes dominate the dynamics.

In order to limit uncertainties in 
the data interpretation we have undertaken a systematic study of the experimental
results as a function of the collision energy and the mass number of
the colliding nuclei.
Sudden changes in the 
studied dependences may indicate that 
a critical collision energy and/or a
critical volume is crossed. 
%
In fact in a previous work we showed that the 
particle multiplicity (entropy) rapidly increases 
in the energy region  between
$\sqrt{s} \approx$ 5 A$\cdot$GeV (BNL AGS) and
$\sqrt{s} \approx$ 20 A$\cdot$GeV (CERN SPS) for large enough colliding
nuclei \cite{Ro:95,Ga:95}.
This increase can be interpreted as due to a transition
to a Quark Gluon Plasma.
Using a statistical approach it was estimated that this transition
is connected with the increase of the effective number of degrees
of freedom by a factor of about 3 \cite{Ga:95}.

In this work we compile the experimental data on strangeness 
production in nuclear collisions.
We study their dependences on the collisions energy and the
volume of the colliding systems.
We attempt to interpret the results within a statistical
approach.
The paper is organized as follows:
The experimental data on strange hadron production in nucleon--nucleon
interactions and central nucleus--nucleus collisions  are
compiled in Sections 2 and 3.
The results are discussed in Section 4 and interpreted in Section 5.
 
\newpage

\section{Nucleon--Nucleon Interactions}

Data on strange particle production in nucleon--nucleon
interactions are necessary as reference data in order to
study strangeness production in
nucleus--nucleus collisions. 
In this section we first compile and discuss data on multiplicity
of  strange hadrons produced in minimum bias
p+p interactions.
Using these data, in the second part of the section,
we construct the strange particle multiplicities
in  nucleon--nucleon interactions.

\subsection{Compilation of p+p Data}

Data on the mean multiplicity of $\Lambda$, $\overline{\Lambda}$
hyperons, $\langle \Lambda \rangle_{pp}$, $\langle \overline{\Lambda} 
\rangle_{pp}$,
and $K^0_S$ mesons, $\langle K^0_S \rangle_{pp}$, produced in minimum bias
p+p interactions \cite{pp} are summarized in Table 1.
The  dependence of $\langle \Lambda \rangle_{pp}$,
$\langle \overline{\Lambda} \rangle_{pp}$ and  
$\langle K^0_S \rangle_{pp}$ on the collision energy is shown in Figs. 1, 2
and 3, respectively.
The multiplicities are plotted against
the Fermi variable \cite{Fe:50}:
\begin{equation}
F = \frac {(\sqrt{s} - 2 \cdot m_p)^{3/4}} {\sqrt{s}^{1/4}},
\end{equation}
where $m_p$ is the proton rest mass and $\sqrt{s}$ is
the collision energy in the center of mass system.
In Figs. 1, 2 and 3 the results obtained at 24.5 GeV/c
\cite{pp} are not included because the $\Lambda$ hyperon multiplicity 
presented there (see Table 1) is two times lower than the value expected
using the systematics of all other measurements.

In this paper $\langle \Lambda \rangle$ denotes the average
multiplicity of $\Lambda$ hyperons 
produced directly (by strong
interactions), originating from electromagnetic decays of $\Sigma^0$
hyperons and in addition a fraction of
$\Lambda$ hyperons from $\Xi$ decays.
Treatment of $\Lambda$ hyperons originating from weak decays
of $\Xi$s is usually not discussed in an explicit way in the
experimental papers. The $\Xi/\Lambda$ ratio is about 3\% at 
$p_{LAB}$ = 19 GeV/c and 10\% at $\sqrt{s}$ = 63 GeV (see
Ref. \cite{Wr:85} and references therein).
From the description of the analysis procedure used in most of
the experiments follows that the $\Lambda$ hyperons from
$\Xi$ decays are (at least in part)
included in the published $\langle \Lambda \rangle$
multiplicity. The same argument applies to the  
$\overline{\Lambda}$ multiplicity.

The data on mean multiplicities 
of $K^+$, $\langle K^+ \rangle_{pp}$,
and $K^-$, $\langle K^- \rangle_{pp}$, mesons produced in minimum bias
p+p interactions \cite{pp} are summarized in Table 2 and shown as a function
of $F$ in Figs. 3 and 4
together with the results for  $\langle K^0_S \rangle_{pp}$.

The compilation is done starting from the threshold proton
incident momentum for strangeness production ($p_{LAB}$ = 2.3 GeV/c)
up to about 400 GeV/c.
This momentum range covers the momentum range of the existing 
nucleus--nucleus data.
The presented data on neutral strange hadron production were
obtained by bubble,  streamer and time projection chamber experiments.
High tracking precision in large fiducial volume is the
characteristic feature of these detectors.
It allows for an efficient measurement of neutral strange particle
decays and their identification by testing the energy--momentum
conservation law at the decay vertex for various decay hypothesis.

The measurement of charged kaon production is done using various
experimental techniques.
At low incident momenta (up to several GeV/c) the charged kaon yields
were extracted by exclusive analysis of hydrogen bubble chamber data
(2.807, 3.67, 4.95, 5.52, 8 and 32 GeV/c \cite{pp}).
At all incident momenta charged kaon production was measured
by counter experiments, where the particle identification
was done by Cherenkov (3.2, 3.7, 12.5, 19.2, 24 and 35--70 GeV/c \cite{pp}) 
and Time--of--Flight (3.349, 3.701, 4.133 GeV/c \cite{pp}) detectors or by 
both methods (ISR) \cite{pp}.
In the case of counter experiments an interpolation between 
points at which differential cross sections were measured is
necessary in order to get the yield of kaons in the full
momentum space.
This interpolation was done in the original papers
for most of the results presented here.

Data on charged kaon production in full momentum space for p+p collisions
at about 11-15 GeV/c (AGS-energies) are sparse.
Therefore, we attempt to evaluate the mean multiplicities of  charged kaons
produced in minimum bias p+p collisions near 14.6 GeV/c from 
spectrometer measurements at
fixed
angles at 12.5 GeV/c and 19.2 GeV/c \cite{pp} which only
partially cover the 
space spanned by rapidity
and transverse momentum. Kaon yields had to be 
extrapolated to the regions
not covered by measurements.
The applied procedure consists of two steps. In the first step 
the measured transverse mass 
spectra were extrapolated to the full transverse mass domain in
order to evaluate the rapidity
density at various rapidities; the details of this
procedure
are described in \cite{Ro:95}.
In the second step
the rapidity densities were interpolated and extrapolated
to the low and high rapidity tails by
a Gaussian  fit. 
The integral of the fitted Gaussian in 
the limits from target rapidity to beam rapidity 
was used as an 
estimation of the mean multiplicity of the charged kaons.
The error on the
multiplicity was taken from the fit procedure and in addition 
an average systematic error stated by the various experiments has been
added linearily to the statistical error. Multiplicity results obtained
by others \cite{Anti,Rossi} using similar extrapolation procedures
are also given.
Various experimental techniques give consistent results
as seen in Table 2 and Figs. 3 and 4.

\subsection{Construction of N+N Multiplicities}

In the following sections we are going to discuss
the strangeness production in central nucleus--nucleus
collisions. Therefore in order to be able to carry out
a meaningful study of the strangeness production dependence
on the collision energy and nucleus mass number, $A$, we have to
construct the results for nucleon--nucleon, N+N, interactions i.e.
for nucleus--nucleus collisions at $A$ = 1 \cite{Ga:91a}.
These results are defined as a weighted average of the data for p+p,
p+n and n+n interactions, where the weight factors are equal 
to the probabilities for finding a given pair of nucleons 
in the projectile and target nuclei.

Unfortunately the results on strange hadron production in p+n
interactions are very poor \cite{pn} and data for n+n interactions
do not exist.
Therefore the construction of the  results  for N+N interactions
has to be based on the p+p data.

Charge symmetry of the strong interaction 
leads to the following relations between strange particle 
multiplicities in proton--proton and neutron--neutron
interactions at the same incident energy \cite{Ga:91a}:
\begin{equation}
\langle \Lambda \rangle_{pp} = \langle \Lambda \rangle_{nn}, 
\end{equation}
\begin{equation}
\langle \overline{\Lambda} \rangle_{pp} = \langle \overline{\Lambda} \rangle_{nn}, 
\end{equation}
\begin{equation}
\langle K^+ \rangle_{pp} = \langle K^0 \rangle_{nn}, 
\end{equation}
\begin{equation}
\langle K^- \rangle_{pp} = \langle \overline{K}^0 \rangle_{nn}, 
\end{equation}
\begin{equation}
\langle K^0 \rangle_{pp} = \langle K^+ \rangle_{nn}, 
\end{equation}
\begin{equation}
\langle \overline{K}^0 \rangle_{pp} = \langle K^- \rangle_{nn}. 
\end{equation}
From the relations for kaon yields (Eqs. 4--7) follow that the mean
multiplicity of produced kaons and antikaons, 
$\langle K + \overline{K} \rangle$, is equal in p+p and n+n interactions.

The strange particle multiplicities in p+n interactions can not be derived 
from p+p results using charge symmetry \cite{Ga:91a}.
Therefore, following the procedure presented in Ref. \cite{Ga:91a},
we assume that:
\begin{equation}
\langle \Lambda \rangle_{pn} = 0.5 \cdot (\langle \Lambda \rangle_{pp} +
\langle \Lambda \rangle_{nn}) =
\langle \Lambda \rangle_{pp},
\end{equation}
\begin{equation}
\langle \overline{\Lambda} \rangle_{pn} = 0.5 \cdot 
(\langle \overline{\Lambda} \rangle_{pp} +
\langle \overline{\Lambda} \rangle_{nn}) =
\langle \overline{\Lambda} \rangle_{pp},
\end{equation}
\begin{equation}
\langle K + \overline{K} \rangle_{pn} = 0.5 \cdot 
(\langle K + \overline{K} \rangle_{pp} +
 \langle K + \overline{K} \rangle_{nn}) =
 \langle K + \overline{K} \rangle_{pp}  .
\end{equation}
The above relations are obeyed within the errors
by the p+n data in the incident momentum range
10--29 GeV/c.

According to the definition of nucleon--nucleon interactions
the mean multiplicity of the particle $a$ in these interactions 
for nucleus--nucleus collisions of identical nuclei is
given by:
\begin{equation}
\langle a \rangle_{NN} =
\left( \frac {Z} {A} \right) ^2 \cdot \langle a \rangle_{pp} +  
2 \cdot \frac {Z} {A} \cdot \left( 1 - \frac {Z} {A} \right)
\cdot \langle a \rangle_{pn} +  
\left( 1 - \frac {Z} {A} \right) ^2 \cdot \langle a \rangle_{nn},  
\end{equation}
where $Z$ and $A$ are the numbers of protons and nucleons in
a nucleus, respectively.
From Eqs. 2--11 follows that:
\begin{equation}
\langle \Lambda \rangle_{NN} = \langle \Lambda \rangle_{pp},
\end{equation}
\begin{equation}
\langle \overline{\Lambda} \rangle_{NN} = \langle \overline{\Lambda} \rangle_{pp},
\end{equation}
\begin{equation}
\langle K + \overline{K} \rangle_{NN} = \langle K + \overline{K} \rangle_{pp},
\end{equation}
independent of $A$ and $Z$ of the colliding systems.

The Eqs. 12--14 are used is the following sections 
to derive the strangeness
yield in nucleon--nucleon interactions.

\section{Central Nucleus--Nucleus Collisions}

In this section we have compiled data on the mean multiplicity of
strange
hadrons produced in central nucleus--nucleus collisions
 at incident momenta of 4.5--200 A$\cdot$GeV/c. 

In the compilation and analysis we included the data on strangeness
production from JINR Synchrophasotron (4.5 A$\cdot$GeV/c),
BNL AGS (11.6 and 14.6 A$\cdot$GeV/c) and 
CERN SPS (200 A$\cdot$GeV/c).
The data on strange particle production below 4.5 A$\cdot$GeV/c
are not included as they are relatively sparse; 
we are not aware of simultaneous measurements of $\Lambda$
hyperons and kaons in the same reaction.

All data on mean multiplicity of $\Lambda$, $\overline{\Lambda}$ hyperons,
$K^0_S$, $K^+$ and $K^-$ mesons produced in central nucleus--nucleus
collisions at energies larger than 4.5 A$\cdot$GeV/c
are presented in Table 3 \cite{AA}.
In the table the corresponding data for p+p interactions at
the same energy per nucleon as the corresponding nucleus--nucleus
results are also given.
If  data on p+p interactions are not
available or of poor quality at the respective incident proton
momentum,
the strange hadron multiplicities were calculated 
by a linear interpolation procedure. 

The data on $\Lambda$ and $K^0_S$ production in central
carbon and oxygen collisions with various nuclear targets
(C--Pb) at 4.5 A$\cdot$GeV/c were obtained by the SKM--200
Collaboration at the Dubna Synchrophasotron \cite{AA}.
The  strange particle decays were detected  by a 2m long streamer
chamber. 
Central collisions (5--15\% of the inelastic cross section)
were defined as those without 
spectator nucleons of the projectile nucleus.
Approximate symmetry of the C+C, C+Ne and O+Ne collisions
allowed for a reliable correction of $K^0_S$ yields for large
acceptance cuts in the target fragmentation region.
The reliability of the correction is lower for asymmetric
collisions and therefore we limit ourself to the analysis
of C+C, C+Ne, O+Ne and C+Cu, C+Zr collisions.
Due to low statistics the data on these collisions
were averaged for the three light systems and for the two heavier.
The data for the light systems are refered in this paper as `O+Ne' 
because the data for O+Ne collisions dominate
the averaged sample.

Charged kaon production in 
central Si+Al
collisions ($7\%$ of the inelastic
cross section) and central Au+Au collisions ($4\%$
of the inelastic cross section) has been
measured by the experiments E802/E859 and E866
\cite{AA} at the AGS.
These experiments published $K^+$ and $K^-$--meson rapidity distributions
covering about 1 to 1.5 rapidity units near midrapidity.
Experiment E810 measured the $\Lambda$ and $K^0_S$ production
in central Si+Si 
collisions (10\% of the inelastic
cross section) \cite{AA}. Here the measurements cover the rapidity region
of $1.4 < y < 3.2$ for $\Lambda$
hyperons  and $2.0 < y < 3.5$ for kaons.
Therefore for all measurements an extrapolation to the 
low and high rapidity tails
was necessary in order to evaluate the mean multiplicity of
strange hadrons in full momentum space. 
In the extrapolation procedure we assumed that the rapidity 
distributions can be parametrized by a Gaussian.
The parameters of the function were fitted to the measured
rapidity distributions. The integral of the fitted 
function  in the limits from target rapidity to beam rapidity  
was used as an estimation of the mean multiplicity. 

The data on mean multiplicities of $\Lambda$, $\overline{\Lambda}$  hyperons,
$K^0_S$, $K^+$ and $K^-$ mesons at CERN SPS (200 A$\cdot$GeV/c)
were obtained by the NA35 Collaboration \cite{AA}.
The data originate mostly from the analysis of strange particle decays
registered by the NA35 streamer chamber. 
Some data on charged kaon production were obtained using the NA35
time projection chamber.
The central collisions (2--6\% of the inelastic cross section)
were defined as those with the
total energy deposited in the projectile spectator region lower
than a given limit.
The large acceptance of the streamer chamber and runs with various
configurations of magnetic field and target position
allowed to obtain strange particle yields in full momentum range
for central S+S and S+Ag collisions \cite{AA}.
The NA36 experiment at CERN SPS also published  data on absolute
strange particle yields in central sulphur--nucleus collisions
at 200 A$\cdot$GeV/c \cite{An:93}. 
Good agreement 
with the NA35 data was shown \cite{An:93}.
In the QM'95 proceedings significant disagreement with NA35 $\Lambda$--data
was reported \cite{Ju:95}, but the results of the revised analysis for 
$\overline{\Lambda}$  and $K^0_S$ were not shown. 
Therefore the NA36 data were not included in the current
compilation.

\section{Strangeness Production in Nuclear Collisions}

In this section we present the basic features of strangeness
production in central nucleus--nucleus collisions at various
energies, its dependence on the sizes of colliding nuclei and
collision energy.
The results are compared with the corresponding data for
nucleon--nucleon interactions.

In order to quantify the total production of strangeness we follow 
a procedure described in Alber et al. \cite{AA} and use
the $E_S$ ratio defined as 
\begin{equation}
E_S = \frac {\langle \Lambda \rangle + \langle K + \overline{K} \rangle}
            {\langle \pi \rangle},
\end{equation}
where $\langle \pi \rangle$ is the mean multiplicity of all pions
produced.
The sum of all kaons and $\Lambda$ hyperons even for the highest energies
discussed here contains more than 70\% of the total number of strange
quarks produced \cite{Wr:85}. 
Since the $\overline{\Lambda}$/$\Lambda$ ratio is always smaller than
$\simeq$ 20\%, the contribution of strange antiquarks in $\overline{\Lambda}$
to $E_S$ is less than 5\%. 
The correction for the unmeasured 
strange particles (remaining main contributors are $\eta$ meson and
$\Sigma^{+/-}$ hyperons) is model--dependent and it was not applied here.
Thus the sum $\langle \Lambda \rangle + \langle K + \overline{K} \rangle$
is assumed in this paper to be proportinal to the total strangeness
production, this assumption may lead to about 10\% systematic biases
when comparing production at different energies and/or for various systems.
The total kaon multiplicity is calculated as:
\begin{equation}
\langle K + \overline{K} \rangle = 
\langle K^+ \rangle + \langle K^- \rangle +
2 \cdot \langle K^0_S \rangle 
\end{equation}
or, if the data on charged or neutral kaons do not exist (see Table 3),
using approximate charge symmetry of the systems as:
\begin{equation}
\langle K + \overline{K} \rangle = 
4 \cdot \langle K^0_S \rangle 
\end{equation}
or
\begin{equation}
\langle K + \overline{K} \rangle = 
2 \cdot (\langle K^+ \rangle + \langle K^- \rangle).
\end{equation}
In the case of central Au+Au collisions at 11.6 A$\cdot$GeV/c only data
on charged kaon multiplicities exist.
Neglecting the small charge asymmetry of the Au+Au system and using
strangeness conservation law we can estimate $\langle \Lambda \rangle$
by:
\begin{equation}
\langle \Lambda \rangle_{AuAu} = 
\frac {2} {1.6 \pm 0.1} \cdot ( \langle K^+ \rangle + \langle K^- \rangle )
= 24 \pm 4,
\end{equation}
where the factor 1.6 is equal to the ratio of the total number of 
produced hyperons to the number of produced $\Lambda$ hyperons
established experimentally for p+p interactions at AGS energies
\cite{Wr:85}.
The procedure was checked using Si+Al/Si data at 14.6 A$\cdot$GeV/c.

In Table 4 we summarize the values of 
$\langle \Lambda \rangle + \langle K + \overline{K} \rangle$,
$\langle \pi \rangle$, the mean number of participant nucleons,
$\langle N_P \rangle$, and the $E_S$ ratio for
central nucleus--nucleus collisions and nucleon--nucleon
interactions at the corresponding energy.
The experimental definition of the number of participant nucleons
used is as follows:
the number of the participant nucleons is equal to the
net baryon number of all particles in the final state 
outside the `Fermi spheres'
(p $<$ 300 MeV/c) of the projectile and target nuclei. 
The description of experimental techniques 
used to measure $\langle N_P \rangle$
can be found in the original papers qouted in the Ref. \cite{Ro:95}.
The $\langle \pi \rangle$ values were calculated 
on the basis of (approximate) charge symmetry of all systems as
\begin{equation}
\langle \pi \rangle = 3\cdot(\langle h^- \rangle - \langle K^- \rangle),
\end{equation}
where $\langle h^- \rangle$ is the mean multiplicity of negatively
charged mesons taken from our previous compilation \cite{Ro:95}.
The $\langle K^- \rangle$ values summarized in the Table 3 are used
for nucleus--nucleus collisions (at 4.5 A$\cdot$GeV/c we assume
$\langle K^- \rangle$ = 0).
For nucleon--nucleon interactions
the $\langle K^- \rangle_{pp}$ values are used (see Table 3) which
may introduce a bias smaller than 1\% \cite{Ga:91a}.
For central Au+Au collisions at 11.6 A$\cdot$GeV/c the recent pion
multiplicity value from Videbaek et al. \cite{AA} was used
with the error deduced by us to be about 10\%.
The $\langle N_P \rangle$ values are taken from the compilation 
\cite{Ro:95}.

The ratio of strangeness per participant nucleon number
defined as 
\begin{equation}
B_S = \frac {\langle \Lambda \rangle + \langle K + \overline{K} \rangle}
            {\langle N_P \rangle}
\end{equation}
is shown in Fig. 5 for central nucleus--nucleus collisions and
nucleon--nucleon interaction as a function of $F$.
The ratio increases for both types of collisions,
being on average a factor of about two larger for nucleus--nucleus
collisions relative to nucleon--nucleon interactions.

The dependence of the $E_S$ ratio (see Eq.~15) on the projectile nucleus
mass number (in all analyzed reactions
the projectile nucleus is smaller than the target one)
at Dubna, AGS and SPS energies is shown in Figs. 6a--6c,
respectively.
Note that the AGS data at 11.6 A$\cdot$GeV/c (Au+Au) 
and 14.6 A$\cdot$GeV/c (Si+Al/Si)
are plotted in the same figure.
The $E_S$ ratio is larger for central nucleus--nucleus collisions
than for nucleon--nucleon interactions at the same energy.
This is observed for all analyzed energies.
The enhancement factor ranges from 1.7 for C+Cu/Zr collisions
at 4.5 A$\cdot$GeV/c and S+S collisions at 200 A$\cdot$GeV/c to
4.5 for Au+Au collisions at 11.6 A$\cdot$GeV/c.
We do not observe any significant dependence of the $E_S$
ratio on the target nucleus mass number at fixed projectile and collision
energy.
This is consistent with the observed independence of the strangeness
suppression factor \cite{Wr:85}, $\lambda_S$,
of the target nucleus mass number for proton--nucleus
interactions \cite{Bi:92}.
The $E_S$ ratio for these interactions is close to the 
$E_S$ ratio for N+N interactions (see e.g. Ref. \cite{Ga:95d}).
It indicates strong projectile dependence, e.g. the ratio
for p+S interactions is about two times lower than the ratio for
S+S collisions at 200 A$\cdot$GeV/c. 
The data at AGS energies suggest a possible saturation  
of the $E_S$ ratio with the volume of the system for collisions
of large enough nuclei.
The increase from Si+Al/Si collisions to Au+Au collisions
is about 40\%, whereas the increase from nucleon--nucleon
interactions to Si+Al/Si collisions is by a factor of about 3.

The energy dependence of the $E_S$ ratio for 
nucleon--nucleon interactions and
central nucleus--nucleus collisions is shown in Fig. 7a and 7b,
respectively.
The $E_S$ ratio increases with the energy in a monotonic way
for nucleon--nucleon interactions (see Fig. 7a).
A qualitatively different behaviour is observed for central
nucleus--nucleus collisions;
the fast increase of the $E_S$ ratio occurs between
Dubna and AGS energies, but the $E_S$ values at AGS and
SPS energies are similar (see Fig. 7b).

\section{Discussion}

We will start the discussion with a possible interpretation
of the presented results in terms of a commonly used approach of
an equilibrated hadronic gas.  In such an approach the freeze-out
temperature is expected to be approximately indpendent of the
collision energy (for high enough energies). This leads to the
weak dependence of $E_S$ on $F$ as indicated by AGS and SPS data.
The rapid increase of $E_S$ between Dubna and AGS energies may be
interpreted as due to an increase of the strangeness saturation level
(towards equilibrium) and/or an increase of the freeze-out 
temperature (possible, providing that the early stage temperature 
is  lower than the high energy limit of the freeze-out 
temperature). This interpretation encountres, however, several problems.
The rapid increase of the pion multiplicity observed between AGS and
SPS energies remains unexplained. A complete description of
the hadronic yields at SPS energies seems to be not possible \cite{Cley,Soll}.
The chemical freeze-out temperature derived from strangeness and baryonic
components is significantly higher at SPS energies ($\approx 200$ MeV 
\cite{Soll}) than at AGS energies ($\approx 110$ MeV \cite{Satz}).
Therefore, an interpretation of the energy dependence of 
pion and strangeness production in central A+A collisions in terms of an
equilibrated hadron gas approach seems to be inappropriate.

A different scenario is presented in the following.
The volume and the collision energy dependence of the entropy production
in nuclear collisions was shown to be consistent with the predictions
of the generalized Landau model \cite{Ga:95,La:53}.
Within this  approach the increase of the entropy 
production  which is observed between AGS and SPS energies \cite{Ro:95}
was interpreted as due to increase of the number of degrees of freedom
in the matter created at the early stage of high energy heavy ion
collisions.
These results motivate the attempt to interpret
data on strangeness production (compiled in this paper) 
within a similar approach.

The presented interpretation is based on the assumption 
that for collisions of large enough nuclei the $E_S$
ratio is proportional to the equilibrium strangeness to entropy
ratio at temperature $T$ which is an increasing function of the 
collision energy\footnote{In the generalized Landau model
the early stage temperature is proportional to $F$ (see Eq. 1)
for $\sqrt{s} >> 2 \cdot m_p$ \cite{Ga:95}.}.

From basic thermodynamics follows that for massive 
particles the equilibrium particle number
to entropy ratio increases with the temperature of the matter.
The ratio reaches its saturation value, equal to  
the corresponding ratio for massless particles,
for $T>>m$.
Thus the increase and a subsequent saturation of the $E_S$
ratio for A+A collisions with $F$ may be interpreted 
as a reflection of the above 
described behaviour, providing that the strangeness carriers
are massive.
This interpretation, however, is inconsistent with the observed rapid
increase of the entropy production occuring between AGS and SPS
energies.

In the following a consistent interpretation is proposed.
The rapid increase of the $E_S$ ratio for A+A collisions 
with $F$ observed at low energies
is due to the mentioned above increase of the equilibrium strangeness
to entropy ratio with $T$ for massive strangeness carriers.
This increase may  continue above the AGS energy.
The $E_S$ ratio for N+N interactions and A+A collisions changes
by a similar factor between Dubna and AGS energies.
Thus the scaled\footnote{The scaling factor is 
equal to the ratio $E_S(Si+Al/Si)/E_S(N+N)$ = 2.8 at 14.6 A$\cdot$GeV/c.}
$E_S$ values for N+N interactions may be used for an approximate extrapolation
of the $E_S$ ratio for A+A collisions above AGS energies
(see open squares and a solid line drawn to guide the eye in Fig. 8).
The $E_S$ increase with $F$ can continue  until
the transition, e.g. from  hadronic or constituent quark matter
\cite{Ch:94}
to a Quark Gluon Plasma, occurs. 
As the equilibrium strangeness to entropy ratio is lower for Quark
Gluon Plasma than for other forms of strongly interacting matter,
the rapid decrease of the $E_S$ ratio is expected in 
the transition region.
Its hypothetical position is between AGS and SPS energies.
The mass of the strangeness carriers is expected to be significantly
lower in the Quark Gluon Plasma ($m_S \approx$ 150 MeV) than in the
hadronic or constituent quark matter ($m_S \approx$ 500 MeV)
due to approximate chiral symmetry restoration.
Therefore the energy dependence of the $E_S$ ratio above the transition 
region should be significantly weaker than below 
the transition\footnote{The quantative results on the temperature
dependence of the strangeness to entropy ratio obtained within simple
models of hadronic matter and Quark Gluon Plasma can be found in
Ref. \cite{Ka:86}.}.
The horizontal dashed line in Fig. 8 indicates the expected $E_S$ dependence
on $F$ in the approximation of  massless strange quarks
($m_S << T$).

\section{Summary and Conclusions}

We would like to summarize the paper as follows.
\begin{itemize}
\item
The data on mean multiplicity of strange hadrons produced in
central nucleus--nucleus collisions and minimum bias proton--proton
interactions at incident momenta 2.8--400 A$\cdot$GeV/c
were compiled.
This allowed the construction of the corresponding results for
nucleon--nucleon interactions which serve as a reference in studying
properties of nucleus--nucleus collisions.
The dependences of the ratios strangeness/pion and strangeness/baryon
on the collision energy and the system size were presented and
discussed.
Strangeness enhancement is observed in central nucleus--nucleus
collisions relative to nucleon--nucleon interactions at all
studied energies.
The ratio strangeness/pion shows qualitatively different behaviour
as a function of collision energy for central nucleus--nucleus
collisions and nucleon--nucleon interactions.
\item
Within a statistical approach  the behaviour of 
the strangeness to pion ratio 
can be interpreted as due to a transition to a new form of matter 
(presumably Quark Gluon Plasma).
This interpretation is consistent with the previous observation
of the transition in the analysis of the entropy production.
%
\end{itemize}
We would like to conclude that experimental data on pion and strangeness 
production in central nucleus--nucleus collisions
in the transition region are needed.

\begin{center}
{\bf Acknowledgments}
\end{center}

We would like to thank J.W. Harris, B. L. Friman, J. Knoll, 
St. Mr\'owczy\'nski and H. Str\"obele
for numerous discussions and critical comments.

\newpage
\begin{center}
{\bf Table Captions}
\end{center}

\noindent
{\bf Table 1} \\ 
\noindent
The compiled results on the mean multiplicities of 
$\Lambda$, $\overline{\Lambda}$ hyperons and $K^0_S$ mesons,
$\langle \Lambda \rangle_{pp}$, $\langle \overline{\Lambda} \rangle_{pp}$ 
and $\langle K^0_S \rangle_{pp}$,
produced in minimum bias proton--proton interactions
at different momenta of incident protons $p_{LAB}$.

\vspace{0.5cm}

\noindent
{\bf Table 2} \\ 
\noindent
The compiled results on the mean  multiplicities of 
charged kaons,
$\langle K^+ \rangle_{pp}$ and $\langle K^- \rangle_{pp}$,
produced in minimum bias proton--proton interactions
at different momenta of incident protons $p_{LAB}$.
In some cases ($^*$) the results of two different extrapolation
procedures are given, 
the symbol denotes the values taken from Ref.~\cite{Anti}. 

\vspace{0.5cm}

\noindent
{\bf Table 3} \\ 
\noindent
The compiled results on the mean  multiplicities of 
strange particles
produced in central nucleus--nucleus collisions
at different momenta per incident nucleon $p_{LAB}$ [A $\cdot$ GeV/c].
The corresponding multiplicities for p+p interactions
obtained by an interpolation procedure are also shown.

\vspace{0.5cm}

\noindent
{\bf Table 4} \\ 
\noindent
The total  mean  multiplicities of 
strange particles, $\langle \Lambda \rangle + \langle K
+ \overline{K} \rangle$, pions, $\langle \pi \rangle$,
participant nucleons, $\langle N_P \rangle$, and
the $E_S$ ratio for central nucleus--nucleus collisions
and nucleon--nucleon interactions
at corresponding incident momenta per nucleon $p_{LAB}$ [ A $\cdot$ GeV/c].

\vspace{0.5cm}

\newpage

{\bf Table 1}

\vspace{0.1cm}

\begin{tabular}{|c|c|c|c|}
\hline
       &         &    &               \cr 
 $p_{LAB}$ [GeV/c]   &~~~~~~~~
~ $ \langle \Lambda \rangle_{pp}$
~~~~~~~~&~~~~~~~~~
~ $ \langle \overline{\Lambda} \rangle_{pp}$
~~~~~~~~&~~~~~~~~~
~ $ \langle K^0_S \rangle_{pp}$
~~~~~~~~~
 \cr
      &           &   &              \cr 
\hline
\hline
    2.807  & 0.00068 $\pm$  0.00019 & -- & --  \cr
    3.67   & 0.0033  $\pm$  0.0006  & -- & 0.00083 $\pm$ 0.00018 \cr
    4.95   & 0.0073  $\pm$  0.0003  & -- & 0.0019  $\pm$ 0.0003  \cr
    5.52   & 0.0127  $\pm$  0.0011  & -- & 0.00364 $\pm$ 0.00010 \cr
    6      & 0.0109  $\pm$  0.0007  & -- & 0.0034  $\pm$ 0.0003  \cr
    6.92   & 0.0172  $\pm$  0.0010  & -- & 0.0064  $\pm$ 0.0005  \cr
    7.87   & 0.0201  $\pm$  0.0010  & -- & 0.0072  $\pm$ 0.0006  \cr
    12     & 0.0388  $\pm$  0.0006  & 0.00010 $^{+ 0.00003}_{- 0.00006}$  &
    0.0202  $\pm$ 0.0004  \cr
    12.4   & 0.035   $\pm$  0.004   & -- & 0.023   $\pm$ 0.003   \cr
    19     & 0.061   $\pm$  0.002   & -- & 0.0415  $\pm$ 0.0013  \cr
    24     & 0.0657  $\pm$  0.0008  & 0.0007 $^{+ 0.0001}_{- 0.0003}$  & 
    0.0495  $\pm$ 0.0006  \cr
    24.5   & 0.037   $\pm$  0.007  & --  &  0.045  $\pm$ 0.004  \cr
    69     & 0.109   $\pm$  0.006   & 0.005  $\pm$ 0.001 &
    0.109   $\pm$ 0.007   \cr
    100    & 0.112   $\pm$  0.013   & 0.008  $\pm$ 0.002 &
    0.122   $\pm$ 0.010   \cr
    102    & 0.099   $\pm$  0.012   & 0.005  $\pm$ 0.002 & 
    0.141   $\pm$ 0.014   \cr
    147    & 0.133   $\pm$  0.006   & 0.025  $\pm$ 0.006 &
    0.158   $\pm$ 0.006   \cr
    200    & 0.08    $\pm$  0.02    & 0.02  $\pm$ 0.01   & 
    0.16    $\pm$ 0.02    \cr
    205    & 0.103   $\pm$  0.011   & 0.012  $\pm$ 0.004 &
    0.181   $\pm$ 0.014   \cr
    300    & 0.111   $\pm$  0.015   & 0.020  $\pm$ 0.004 &
    0.21    $\pm$ 0.03    \cr
    300    & 0.11    $\pm$  0.01    & 0.031  $\pm$ 0.009 & 
    0.224   $\pm$ 0.018   \cr
    360    & 0.12    $\pm$  0.02    & 0.013  $\pm$ 0.004 & 
    0.26    $\pm$ 0.01    \cr
    400    & 0.12    $\pm$  0.01    & 0.013  $\pm$ 0.003 &
    0.20    $\pm$ 0.02    \cr
    405    & 0.125   $\pm$  0.008   & 0.020  $\pm$ 0.004 & 
    0.232   $\pm$ 0.011   \cr
\hline
\end{tabular}
\newpage

{\bf Table 2}

\vspace{0.1cm}

\begin{tabular}{|c|c|c|}
\hline
       &         &                   \cr 
 $p_{LAB}$ [GeV/c]   &~~~~~~~~
~ $ \langle K^+ \rangle_{pp}$
~~~~~~~~&~~~~~~~~~
~ $ \langle K^- \rangle_{pp}$
~~~~~~~~~
 \cr
      &           &                 \cr 
\hline
\hline
   2.807  & 0.00068  $\pm$ 0.00019  & --  \cr
   3.2    & 0.0019   $\pm$ 0.0003~$^*$   & --  \cr
   3.2    & 0.0026   $\pm$ 0.0007   & --  \cr
   3.349  & 0.0022   $\pm$ 0.0004   & --  \cr
   3.67   & 0.0046   $\pm$ 0.0007   & --  \cr
   3.7    & 0.0033   $\pm$ 0.0005~$^*$   & --  \cr
   3.7    & 0.0046   $\pm$ 0.0011   & --  \cr
   3.701  & 0.0044   $\pm$ 0.0008   & --  \cr
   4.133  & 0.0057   $\pm$ 0.0010   & --  \cr
   4.95   & 0.0069   $\pm$ 0.0010   & --  \cr
   5.52   & 0.0080   $\pm$ 0.0017   & --  \cr
   8      & 0.020    $\pm$ 0.004    & --  \cr
   12.5   & 0.07     $\pm$ 0.03     & 0.010 $\pm$ 0.004  \cr
   12.5   & --                      & 0.008 $\pm$ 0.001~$^*$  \cr
   14.3   & 0.054    $\pm$ 0.008    & --  \cr
   19.2   & 0.107    $\pm$ 0.016    & 0.036 $\pm$ 0.005  \cr
   24     & 0.102    $\pm$ 0.01     & 0.027 $\pm$ 0.003  \cr
   24     & 0.13     $\pm$ 0.02~$^*$    & 0.033 $\pm$ 0.005~$^*$  \cr
   32     & --                      & 0.05  $\pm$ 0.015  \cr
   35     & --                      & 0.07  $\pm$ 0.02   \cr
   43     & --                      & 0.08  $\pm$ 0.02   \cr
   52     & --                      & 0.11  $\pm$ 0.03   \cr
   70     & 0.21     $\pm$ 0.06     & 0.13  $\pm$ 0.03   \cr
  289     & 0.337    $\pm$ 0.051    & 0.209 $\pm$ 0.031  \cr
  498     & 0.367    $\pm$ 0.055    & 0.244 $\pm$ 0.037  \cr
\hline
\end{tabular}
\newpage

{\bf Table 3}

\vspace{0.1cm}

\begin{tabular}{|c|c|c|c|c|c|c|}
\hline
       &   & & &      &       &            \cr 
 $p_{LAB}$  &Reaction& $\langle \Lambda \rangle$ 
& $\langle \overline{\Lambda} \rangle$
& $\langle K^0_S \rangle$ &
$ \langle K^+ \rangle$ &
$ \langle K^- \rangle$ \cr
      &    & & &       &      &           \cr 
\hline
\hline
4.5& p+p   & 0.006$\pm$0.001 & -- & 0.0019$\pm$0.0004 & 
              0.006$\pm$0.001 & 0 \cr
4.5&`O+Ne' & 0.091$\pm$0.017 & $\le$ 0.001 & 0.035$\pm$0.014 & - & -  \cr
4.5&C+Cu/Zr& 0.112$\pm$0.018 & $\le$ 0.001 & 0.048$\pm$0.016 & - & -  \cr
\hline
11.6& p+p & 0.038$\pm$0.001 & (0.9$\pm$0.6)10$^{-4}$ & 0.019$\pm$ 0.002 & 
             0.051$\pm$0.017 & 0.010$\pm$0.005  \cr
11.6&Au+Au& -- &  -- & -- & 24$\pm$2 & 5.0$\pm$0.4  \cr
\hline
14.6& p+p & 0.047$\pm$0.002 &  (2.1$\pm$1.5)10$^{-4}$ & 0.027$\pm$0.003 & 
             0.065$\pm$0.017 & 0.015$\pm$0.007 \cr
14.6&Si+Al/Si& 2.4$\pm$0.2 & -- & 1.47$\pm$0.06 & 2.6$\pm$0.3 & 0.66$\pm$0.08 \cr
\hline
200& p+p & 0.096$\pm$0.01 & 0.013$\pm$0.005 &0.17 $\pm$ 0.01& 0.28$\pm$0.06&0.18$\pm$0.05\cr
200& S+S & 9.4  $\pm$ 1.0 & 2.2$\pm$0.4 & 10.5 $\pm$ 1.7 & 12.5$\pm$0.4 & 6.9$\pm$0.4 \cr
200& S+Ag& 15.2 $\pm$ 1.2 & 2.6$\pm$0.3 & 15.5 $\pm$ 1.5 & 17.4$\pm$1.0 & 9.6$\pm$1.0 \cr
\hline
\end{tabular}
\newpage

{\bf Table 4}

\vspace{0.1cm}

\begin{tabular}{|c|c|c|c|c|c|}
\hline
       &   & & &      &                   \cr 
 $p_{LAB}$  & Reaction & $\langle \Lambda \rangle + 
\langle K + \overline{K} \rangle$ 
& $\langle \pi \rangle$ &
$ \langle N_P \rangle$ &
 $E_S$ \cr
      &    & & &       &                 \cr 
\hline
\hline
4.5 & N+N   & 0.0158  $\pm$ 0.0015 & 1.87 $\pm$ 0.08 & 
              2 & 0.0084$\pm$0.0009 \cr
4.5 &`O+Ne' & 0.23 $\pm$ 0.06 & 13.2 $\pm$ 0.6 & 19$\pm$2 & 0.017$\pm$0.005  \cr
4.5 &C+Cu/Zr& 0.30 $\pm$ 0.07 & 21.6 $\pm$ 0.9 & -- & 0.014$\pm$0.003 \cr
\hline
11.6 & N+N & 0.137 $\pm$ 0.018 & 3.12 $\pm$ 0.08 & 
             2 & 0.044$\pm$0.006  \cr
11.6 &Au+Au& 82 $\pm$ 6 & 420 $\pm$ 50 & 355 $\pm$ 7  & 0.20$\pm$0.03\cr
\hline
14.6 & N+N & 0.181 $\pm$ 0.19 & 3.44 $\pm$ 0.08 & 
             2 & 0.053$\pm$0.006 \cr
14.6 &Si+Al/Si& 8.6 $\pm$ 0.4 & 58 $\pm$ 6 & 42 $\pm$ 4 & 0.148$\pm$0.017 \cr
\hline
200 & N+N & 0.896 $\pm$ 0.08 &  9.15 $\pm$ 0.08 & 2 & 0.098$\pm$0.009 \cr
200 & S+S & 50    $\pm$ 4 & 273 $\pm$ 10 & 51.2 $\pm$ 5 & 0.183$\pm$0.016 \cr
200 & S+Ag& 77 $\pm$ 6 & 474 $\pm$ 30 & 90 $\pm$ 10 & 0.162$\pm$0.016 \cr
\hline
\end{tabular}
\newpage

\begin{center}
{\bf Figure Captions}
\end{center}

\noindent
{\bf Fig. 1.} \\ 
\noindent
Dependence of the mean multiplicity of $\Lambda$ hyperons,
$\langle \Lambda \rangle_{pp}$, produced in minimum bias 
proton--proton interactions on the Fermi variable
$F$ (see Eq. 1).

\vspace{0.5cm}

\noindent
{\bf Fig. 2.} \\ 
\noindent
Dependence of the mean multiplicity of $\overline{\Lambda}$ hyperons,
$\langle \overline{\Lambda} \rangle_{pp}$, produced in minimum bias 
proton--proton interactions on the Fermi variable
$F$ (see Eq. 1).

\vspace{0.5cm}

\noindent
{\bf Fig. 3.} \\ 
\noindent
Dependence of the mean multiplicity of $K^0_S$ mesons,
$\langle K^0_S \rangle_{pp}$ (circles), $K^+$ mesons,
$\langle K^+ \rangle_{pp}$ (squares), and $K^-$ mesons,
$\langle K^- \rangle_{pp}$ (triangles),  
produced in minimum bias 
proton--proton interactions on the Fermi variable
$F$ (see Eq. 1).

\vspace{0.5cm}

%
%
%
%
%

\noindent
{\bf Fig. 4.} \\ 
\noindent
Dependence of the mean multiplicity of $K^0_S$ (circles),
$K^+$ (squares) and $K^-$ (triangles) 
mesons produced in minimum bias 
proton--proton interactions near the threshold on the Fermi variable
$F$ (see Eq. 1).

\vspace{0.5cm}

\noindent
{\bf Fig. 5.} \\ 
\noindent
Dependence of the $B_S$ ratio (see Eq. 21)
on the Fermi variable
$F$ (see Eq. 1) for nucleon--nucleon interactions (squares) and
central nucleus--nucleus collisions (circles).

\vspace{0.5cm}

\noindent
{\bf Fig. 6.} \\ 
\noindent
Dependence of the $E_S$ ratio (see Eq. 15)
on the projectile  
nucleus mass number, $A_P$,
at 4.5 A$\cdot$GeV/c (a), 11.6 and 14.6 A$\cdot$GeV/c (b)
and 200 A$\cdot$GeV/c (c).
Two nucleon--nucleon points in Fig. (b) correspond to  
11.6 and 14.6 A$\cdot$GeV/c.
\vspace{0.5cm}

\noindent
{\bf Fig. 7.} \\ 
\noindent
Dependence of the $E_S$ ratio (see Eq. 15)
on the Fermi variable
$F$ (see Eq. 1) for nucleon--nucleon interactions (a) and
central nucleus--nucleus collisions (b).

\vspace{0.5cm}

\noindent
{\bf Fig. 8.} \\ 
\noindent
Dependence of the $E_S$ ratio (see Eq. 15)
on the Fermi variable
$F$ (see Eq. 1) for scaled (see text)
nucleon--nucleon interactions (open squares) and
central nucleus--nucleus collisions (circles).
See text for explanation of the lines.


\vspace{0.5cm}

\pagebreak
\begin{figure}[hbt]
\centerline{\epsfig{figure=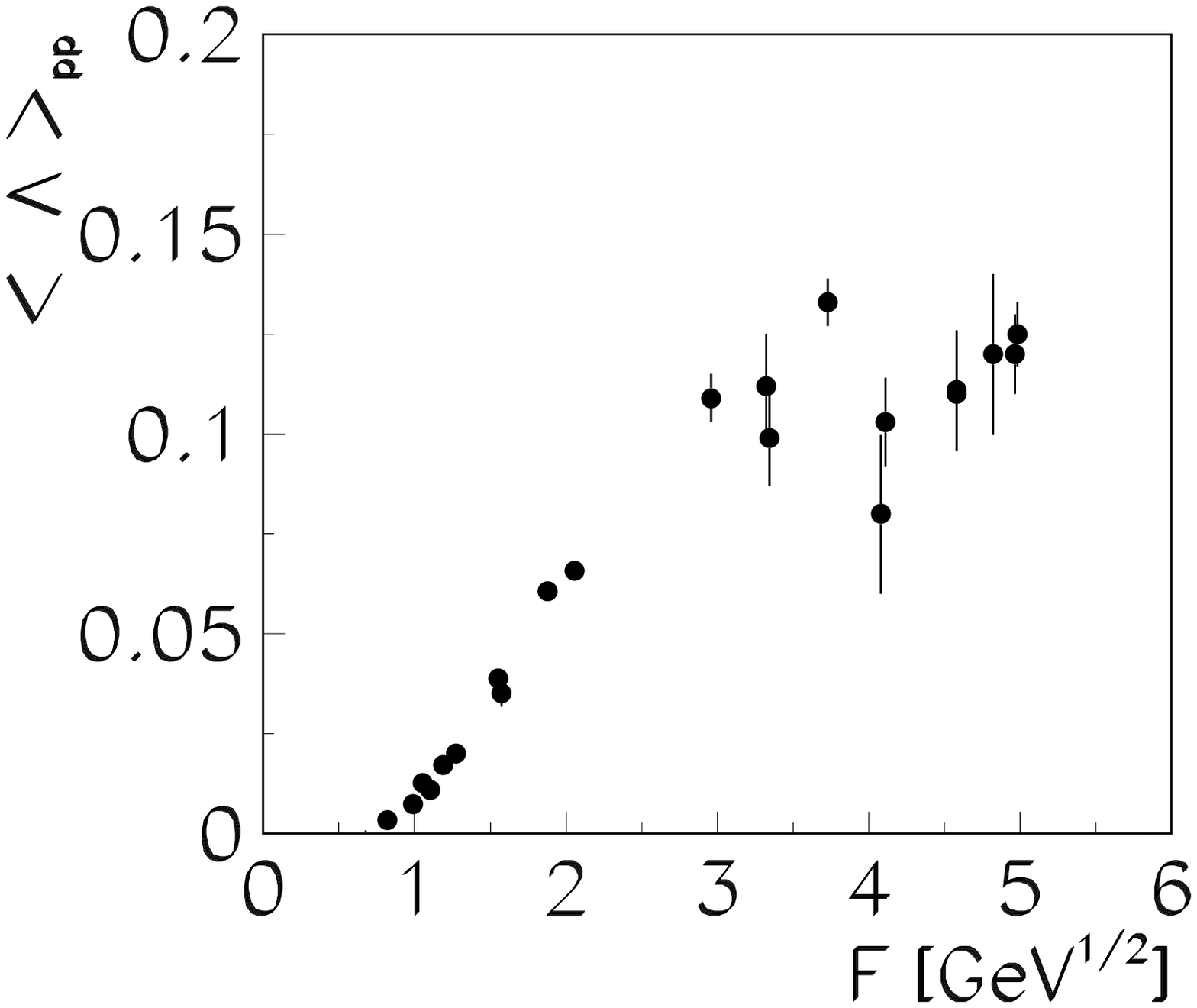,height=14cm,width=16cm}}
\caption{ }
\end{figure}
\pagebreak

\begin{figure}[hbt]
\centerline{\epsfig{figure=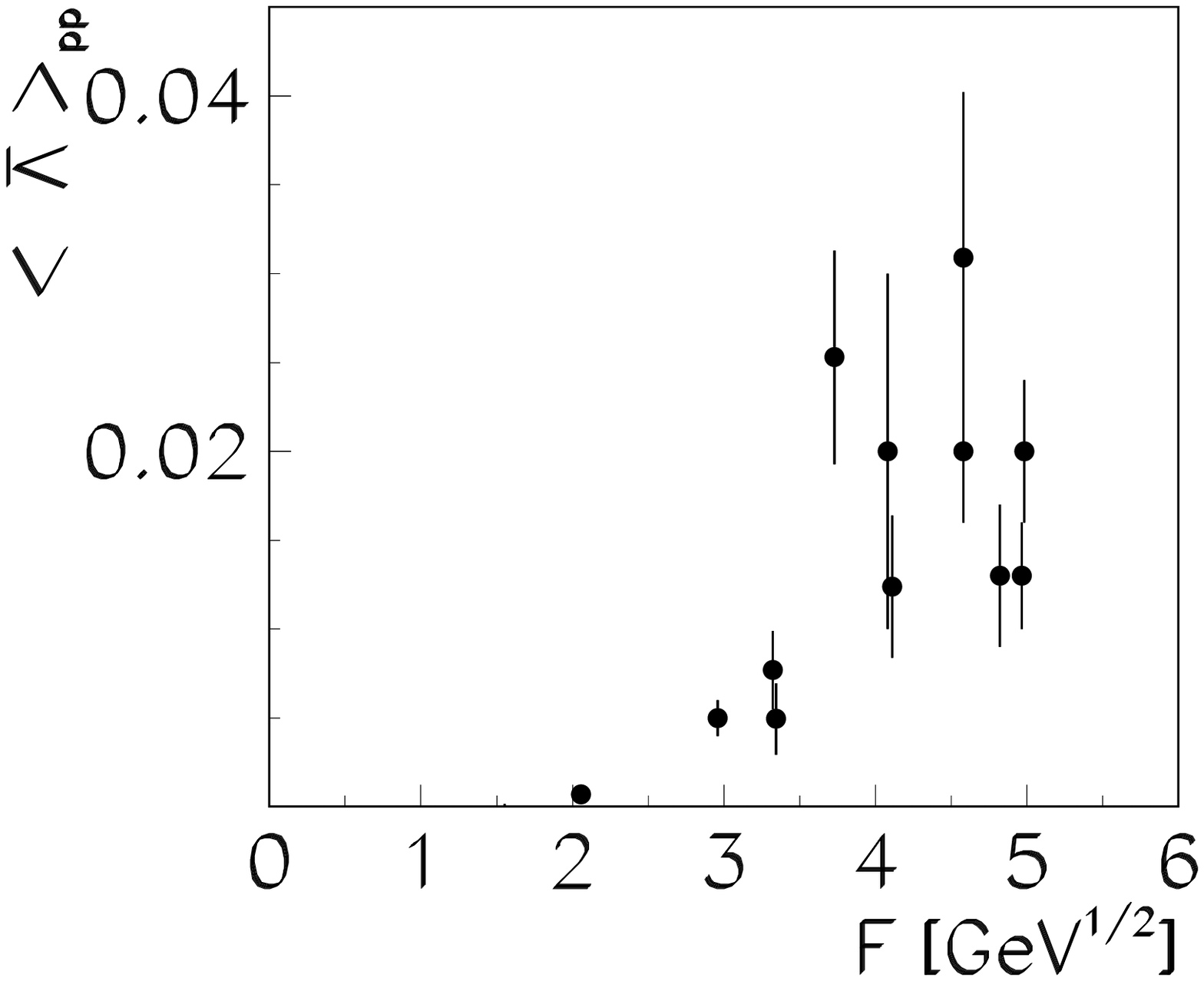,height=14cm,width=16cm}}
\caption{ }
\end{figure}
\pagebreak

\begin{figure}[hbt]
\centerline{\epsfig{figure=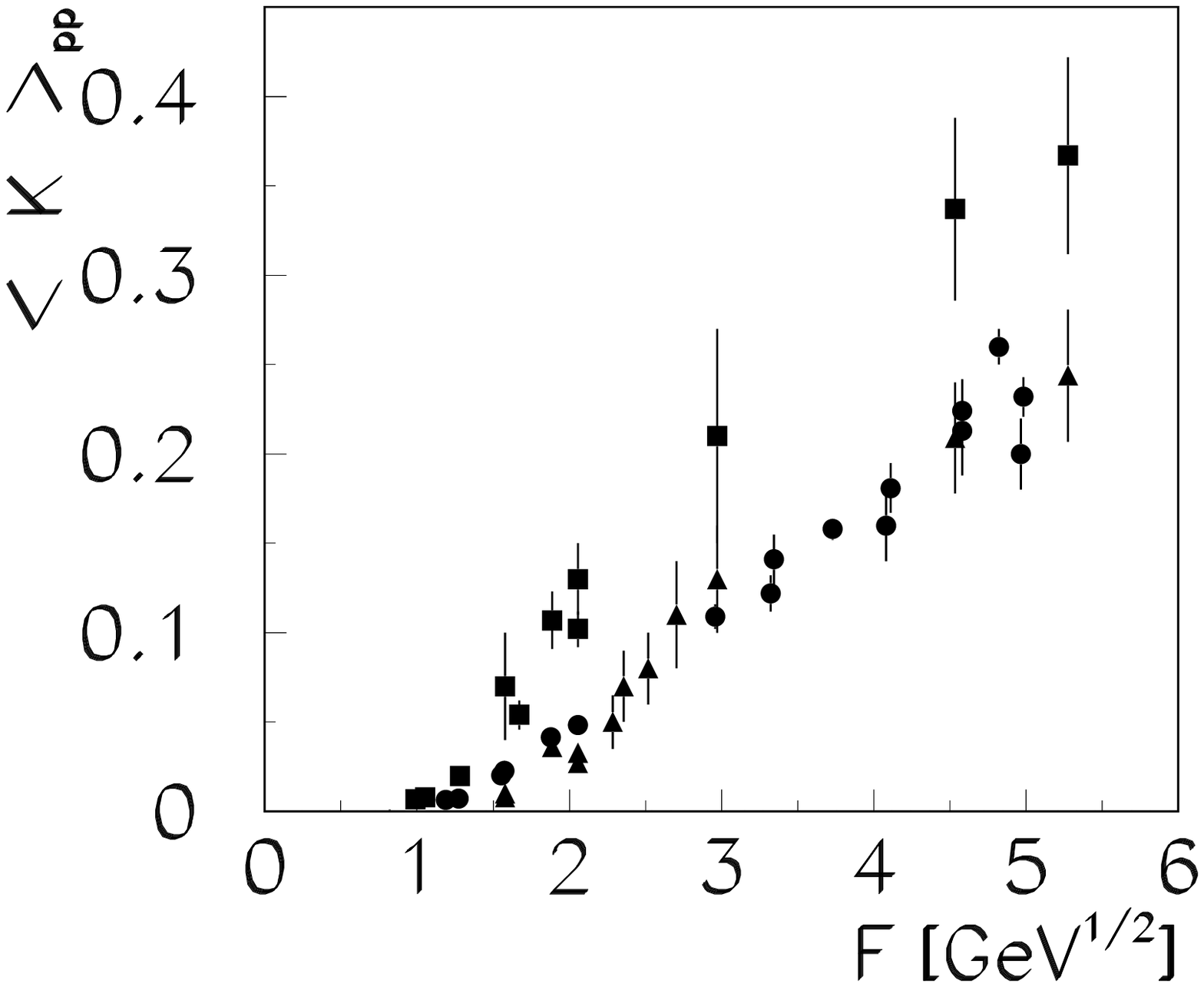,height=14cm,width=16cm}}
\caption{ }
\end{figure}
\pagebreak

%
%
%
%
%
%
%
%
%
\begin{figure}[hbt]
\centerline{\epsfig{figure=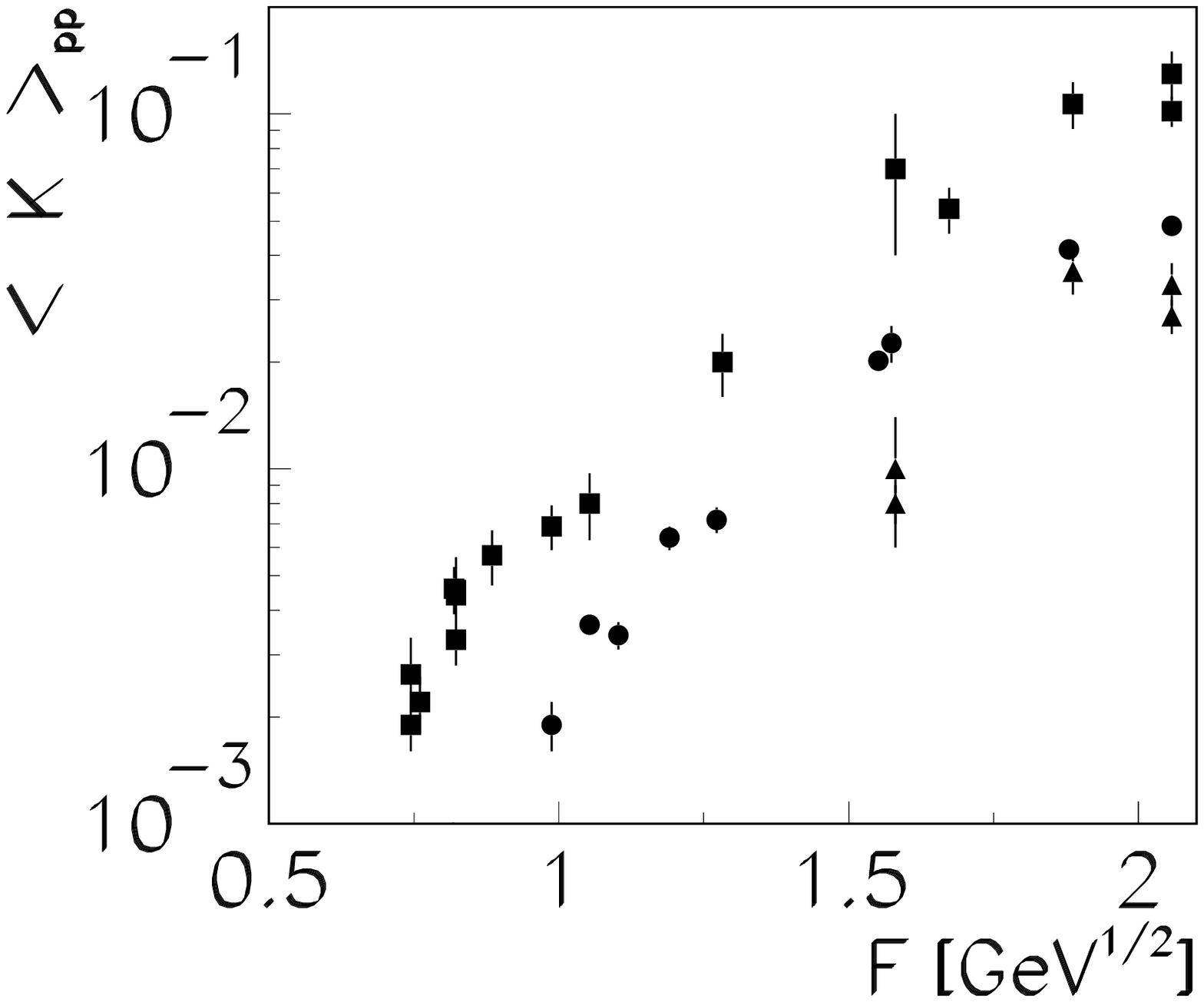,height=14cm,width=16cm}}
\caption{ }
\end{figure}
\pagebreak
\begin{figure}[hbt]
\centerline{\epsfig{figure=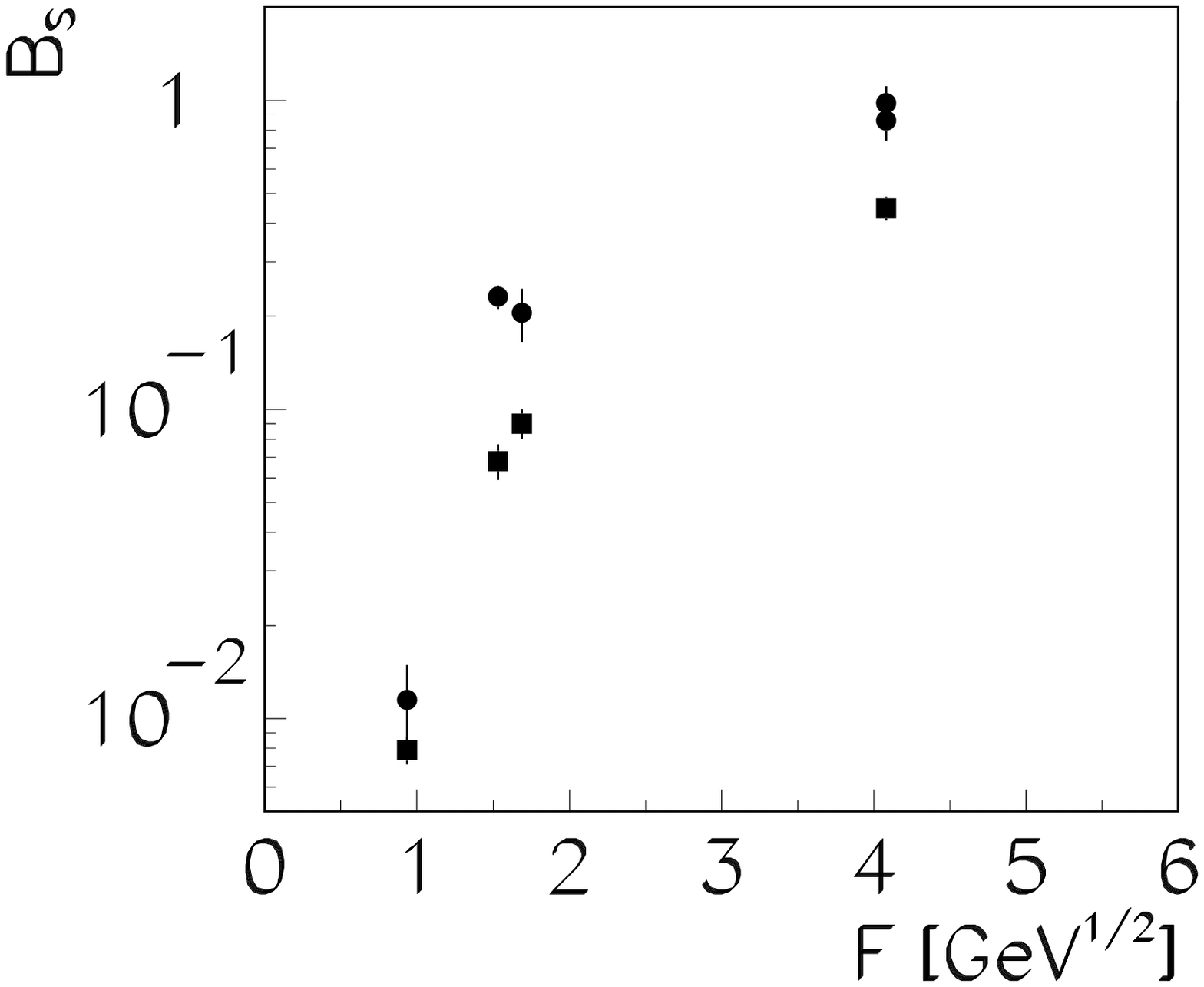,height=14cm,width=16cm}}
\caption{ }
\end{figure}
\pagebreak
\begin{figure}[hbt]
\centerline{\epsfig{figure=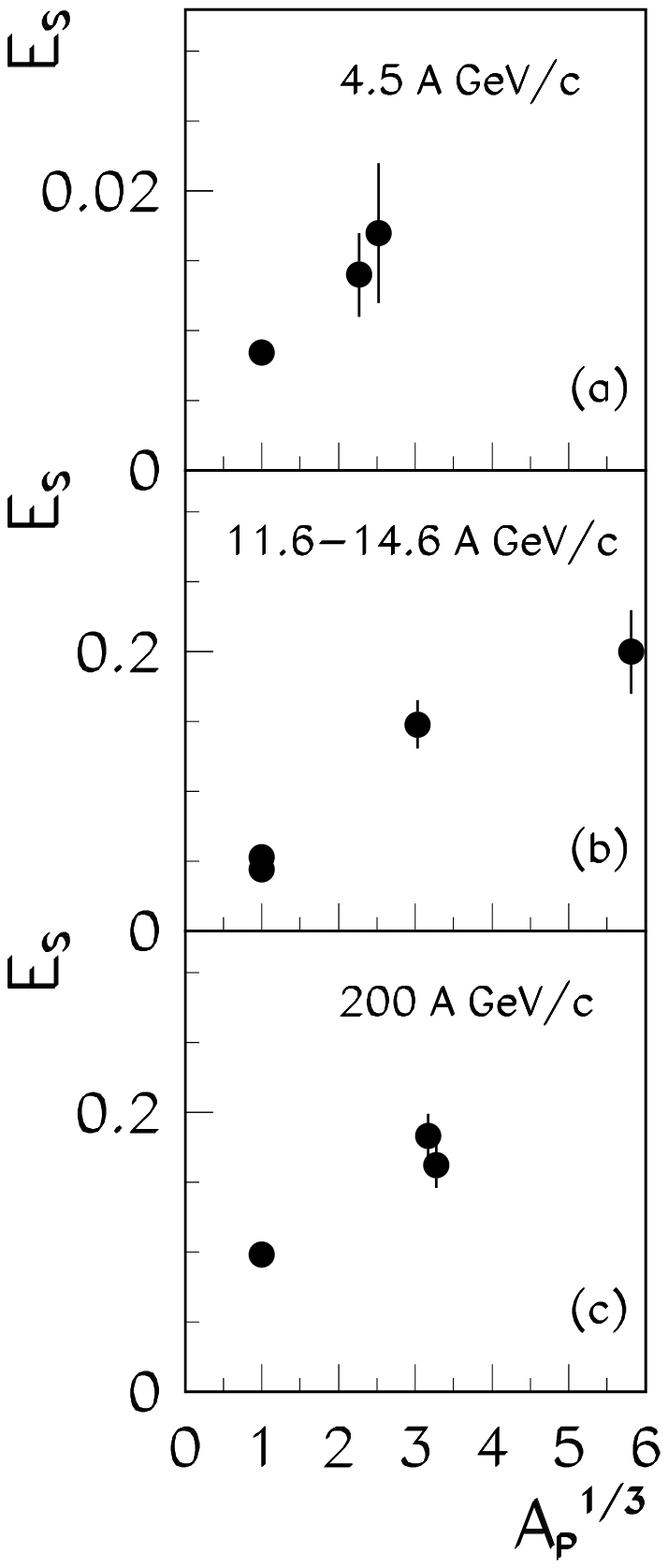,height=20cm,width=12cm}}
\caption{ }
\end{figure}
\pagebreak
\begin{figure}[hbt]
\centerline{\epsfig{figure=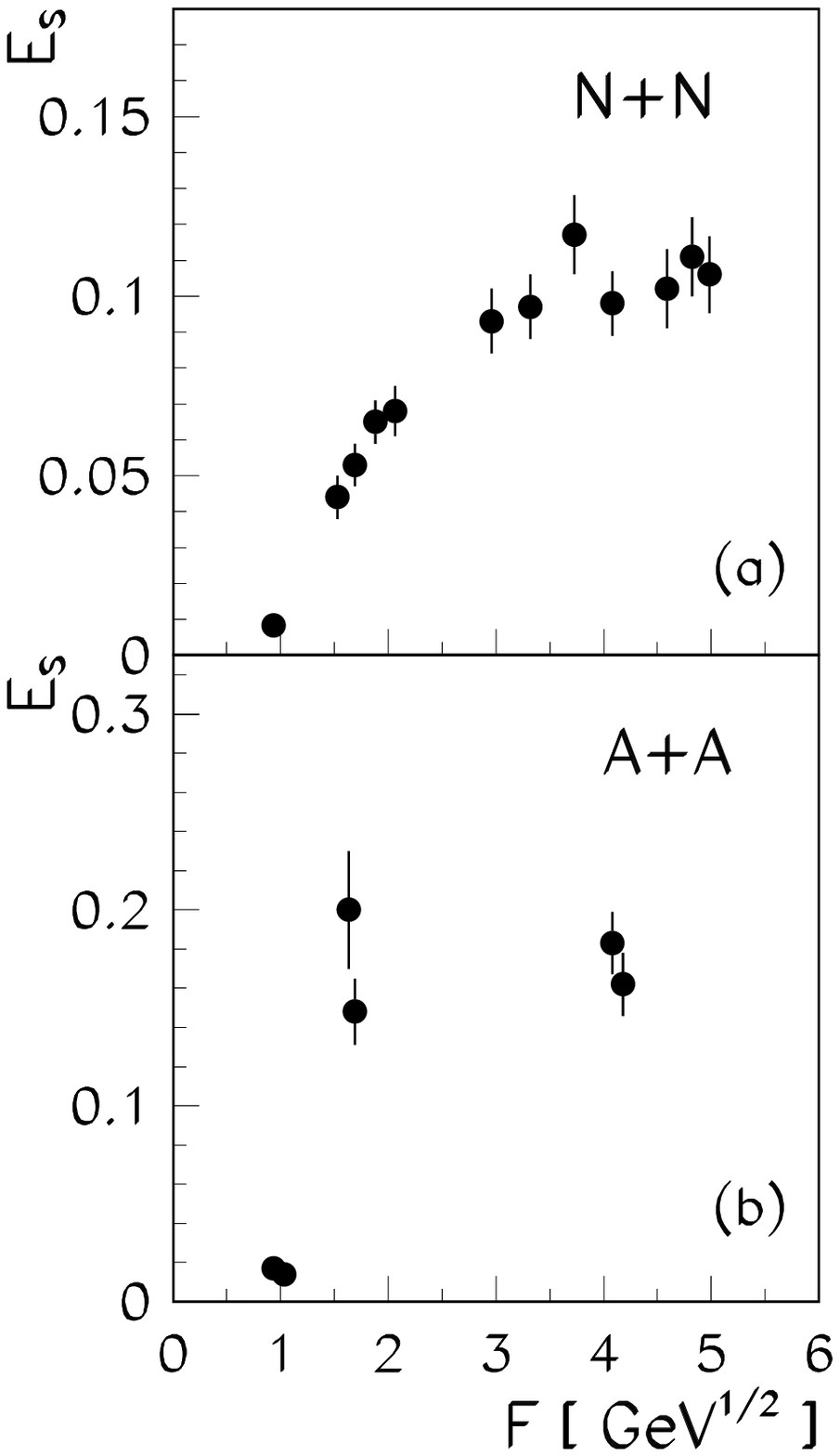,height=18cm,width=13cm}}
\caption{ }
\end{figure}
\pagebreak
%
%
\pagebreak
\begin{figure}[hbt]
\centerline{\epsfig{figure=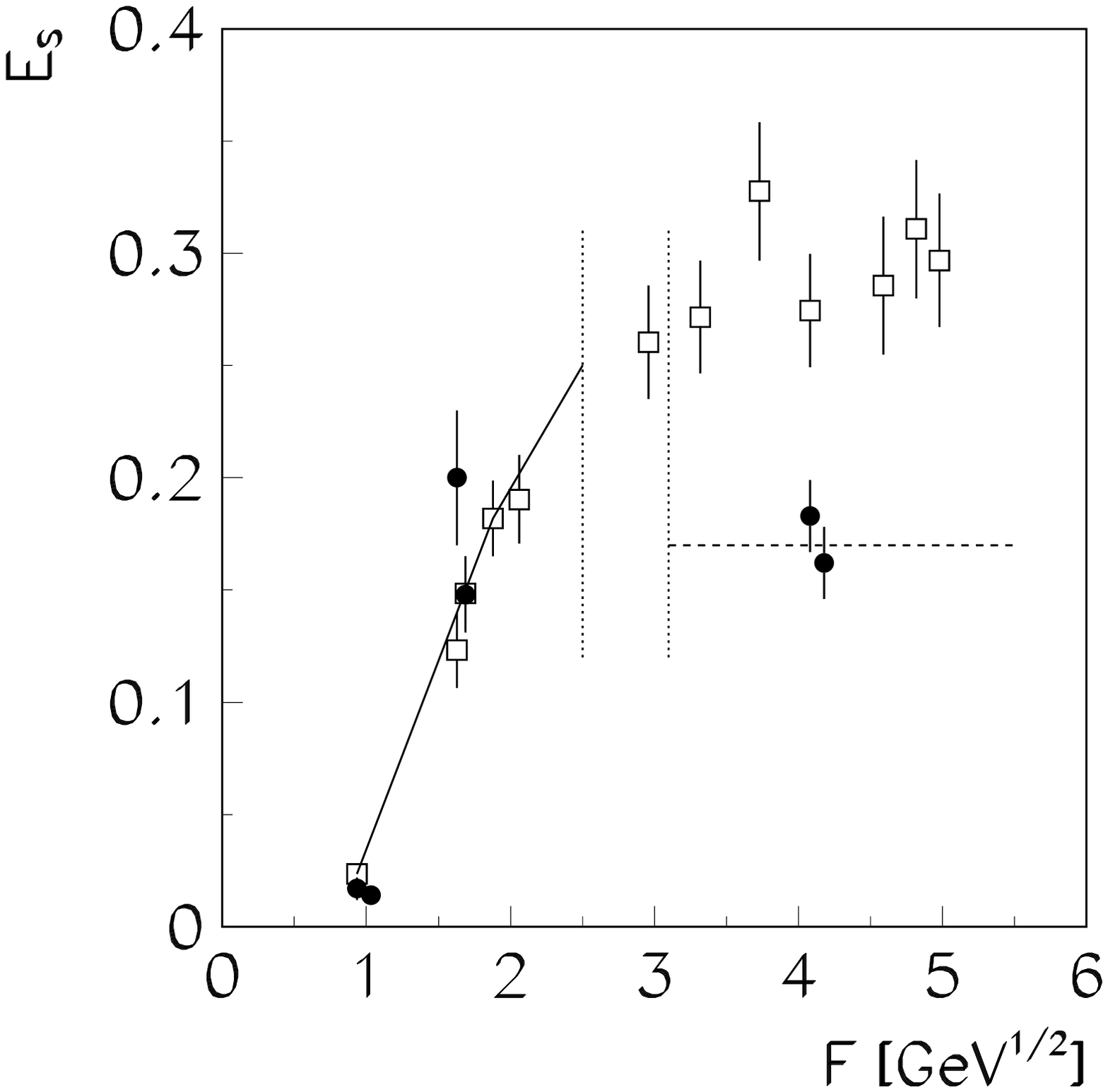,height=14cm,width=16cm}}
\caption{ }
\end{figure}
\pagebreak
\end{document}